\documentclass[10pt, final, conference, letterpaper]{IEEEtran}
\usepackage[left=.6in,right=.6in,top=.719in,bottom=.99in]{geometry}

\usepackage{amsmath,amssymb,dsfont,stfloats,color,url,bbm}
\usepackage[pdftex]{graphicx}
\usepackage[caption=false,font=footnotesize]{subfig}
\usepackage{mathabx}
\usepackage{multirow}
\usepackage{tabu}

\usepackage{siunitx}
\usepackage{tikz} 
\usepackage{pgfplots}
\pgfplotsset{compat=newest} 
\usepgfplotslibrary{units} 
\DeclareGraphicsExtensions{.eps,.pdf,.png,.jpg,.gif,.jpeg,.pstex}

\usepackage[noadjust]{cite}
\usepackage{url}

\newfont{\bbb}{msbm10 scaled 700}

\newfont{\bb}{msbm10 scaled 1100}
\newcommand{\CC}{\mbox{\bb C}}
\newcommand{\PP}{\mbox{\bb P}}

\newcommand{\EE}{\mbox{\bb E}}

\newcommand{\HH}{\mbox{\bb H}}

\newcommand{\yy}{\mathbbm{y}}

\newcommand{\zz}{\mathbbm{z}}
\newcommand{\sss}{\mathbbm{s}}

\newcommand{\hh}{\mathbbm{h}}

\newcommand{\vvv}{\mathbbm{v}}

\usepackage[mathscr]{euscript}
\newcommand{\Rs}{\mathscr{R}}

\newcommand{\av}{{\bf a}}

\newcommand{\hv}{{\bf h}}

\newcommand{\vv}{{\bf v}}
\newcommand{\xv}{{\bf x}}

\newcommand{\zerov}{{\bf 0}}

\newcommand{\Fm}{{\bf F}}

\newcommand{\Rm}{{\bf R}}

\newcommand{\Ac}{{\cal A}}

\newcommand{\Cc}{{\cal C}}

\newcommand{\Ec}{{\cal E}}

\newcommand{\Gc}{{\cal G}}

\newcommand{\Kc}{{\cal K}}

\newcommand{\Nc}{{\cal N}}

\newcommand{\Rc}{{\cal R}}
\newcommand{\Sc}{{\cal S}}

\newcommand{\Uc}{{\cal U}}

\newcommand{\nuv}{\hbox{\boldmath$\nu$}}

\renewcommand{\arg}{{\hbox{arg}}}

\newcommand{\eqdef}{\stackrel{\Delta}{=}}
\newcommand{\defines}{{\,\,\stackrel{\scriptscriptstyle \bigtriangleup}{=}\,\,}}

\newcommand{\herm}{{\sf H}}

\newcommand{\SINR}{{\sf SINR}}
\newcommand{\SNR}{{\sf SNR}}

\newcommand{\taudmrs}{\tau_p}

\newcommand{\Ktot}{K} 
\newcommand{\Kact}{K_{\rm act}}
\newcommand{\Kactset}{\Kc_{\rm act}}

\usepackage{stmaryrd} 

\renewcommand{\arg}{{\rm arg}}

\begin{document}

\setlength{\abovedisplayskip}{1pt}
\setlength{\belowdisplayskip}{1pt}
\setlength{\abovedisplayshortskip}{1pt}
\setlength{\belowdisplayshortskip}{1pt}

\title{Fairness Scheduling in Dense User-Centric Cell-Free Massive MIMO Networks}

\author{\IEEEauthorblockN{Fabian G\"ottsch\IEEEauthorrefmark{1},
		Noboru Osawa\IEEEauthorrefmark{2}, Takeo Ohseki\IEEEauthorrefmark{2}, Yoshiaki Amano\IEEEauthorrefmark{2}, Issei Kanno\IEEEauthorrefmark{2}, Kosuke Yamazaki\IEEEauthorrefmark{2}, Giuseppe Caire\IEEEauthorrefmark{1}}
	\IEEEauthorblockA{\IEEEauthorrefmark{1}Technical University of Berlin, Germany\\
		\IEEEauthorrefmark{2}KDDI Research Inc., Japan\\
		Emails: \{fabian.goettsch, caire\}@tu-berlin.de, \{nb-oosawa, ohseki, yo-amano, is-kanno, ko-yamazaki\}@kddi-research.jp}}

\maketitle


\begin{abstract}
We consider a user-centric scalable cell-free massive MIMO network with a total of $LM$ distributed remote radio unit antennas serving $K$ user equipments (UEs). Many works in the current literature assume $LM\gg K$, enabling high UE data rates but also leading to a system not operating at its maximum performance in terms of sum throughput. We provide a new perspective on cell-free massive MIMO networks, investigating rate allocation and the UE density regime in which the network makes use of its full capability. The UE density $K$ approximately equal to $\frac{LM}{2}$ is the range in which the system reaches the largest sum throughput. In addition, there is a significant fraction of UEs with relatively low throughput, when serving $K>\frac{LM}{2}$ UEs simultaneously. We propose to reduce the number of active UEs per time slot, such that the system does not operate at ``full load'', and impose throughput fairness among all users via a scheduler designed to maximize a suitably defined concave componentwise non-decreasing network utility function. Our numerical simulations show that we can tune the system such that a desired distribution of the UE throughput, depending on the utility function, is achieved.
\end{abstract}

\section{Introduction} 
Based on multiuser MIMO and Marzetta’s massive MIMO \cite{caire2003achievable, 3gpp38211, marzetta2010noncooperative}, distributed MIMO architectures have been promoted throughout the last years for beyond 5G networks to serve the growing number of wireless devices (see \cite{9336188} for an overview).
User-centric scalable cell-free massive MIMO is a branch of distributed MIMO, where each user equipment (UE) $k\in [\Ktot]$\footnote{We denote the set of the first positive $N$ integers by $[N] = \{ 1, \dots, N \}$.} is associated to its own finite size set $\Cc_k$ of remote radio units (RUs), and each RU $\ell \in [L]$, equipped with $M$ antennas, serves a finite size set $\Uc_\ell$ of UEs. 
Many works on cell-free massive MIMO consider $LM \gg \Ktot$, i.e., there are many more RU antennas than UEs being served on the same time-frequency resource. This needs to be carefully discussed since 1) in this regime the network is not working at full capability and 2) the deployment of more RU antennas than UEs being served by the network is not practical, costly, and often infeasible. A popular system performance metric is the distribution of the achievable ergodic user rates, but fairness is often not addressed as an optimization goal. In this work, we investigate the optimal user load at which the maximum sum throughput is achieved, where the throughput is the long-term average service rate considering information outage. When users are scheduled on a slot-by slot basis, rate allocation is necessary and an important factor for the system performance. Ergodic rates are thus not as meaningful, so that we consider the sum throughput under rate allocation and information outage. The maximum sum throughput is achieved with $K \approx \frac{LM}{2}$, where a user load $K > \frac{LM}{2}$ results in a non-negligible number of UEs suffering from relatively low rates.

Therefore, we propose to keep the network at a user load for which the network does not operate at ``full load'', i.e., the number of active UEs on any time slot is chosen such that each UE is expected to experience a relatively large data rate.
Since not all users are active in each time slot, we need to impose some form of throughput fairness. The actual supported instantaneous user service rate is a random variable and we operate in the outage rate regime, which means that an allocated transmission rate may or may not correspond to a successful transmission. 
Thus, the scheduler must select the set of active users and their transmission rates, as these depend on the active user itself.

Scheduling in cell-free massive MIMO has been considered in a number of other works, among others \cite{chen2019dynamic, ammar2021distributed, denis2021improving, ming2022downlink}. User scheduling in co-located and cell-free massive MIMO systems using the framework of \cite{georgiadis2006resource} is studied in \cite{chen2019dynamic}, where all UEs are connected to all RUs. Beamforming with power allocation and user scheduling in a distributed manner based on a signal to leakage, interference and noise ratio metric is done in \cite{ammar2021distributed}. 
In this paper, we start with an investigation of the optimal user load per time slot for a \textit{dense user-centric scalable} cell-free network, where $\Kact$ is approximately the optimal number of simultaneously active UEs. The scheduler is thus designed to select $\Kact$ simultaneously active UEs per time slot.
Compared to previous works, this is the first work on scheduling in cell-free massive MIMO considering rate allocation under slot-by-slot coding and decoding, i.e., the receiver can only decode a rate provided that no information outage occurs. 
In addition, we propose an online scheme to learn the allocated rates, aiming to maximize the expected service rates. The expected service rates are given by the product of the expected data rate to be provided to a UE and the probability that the data rate can be provided without information outage. 
We use the framework of \cite{georgiadis2006resource} to develop a hard fairness scheduler (HFS) and a proportional fairness scheduler (PFS). The results show that the schedulers can improve fairness among users compared to random, round-robin and max-sum-rate scheduling. 

\section{System description} 

We refer the reader to our previous work \cite{goettsch2021impact} for a detailed system model description, and provide a summary in the following. We consider a cell-free wireless network in TDD operation mode with $L$ RUs, each equipped with $M$ antennas, 
and $\Ktot$ single-antenna UEs. Both RUs and UEs are distributed on a squared region on the 2-dimensional plane. 
As a result of the cluster formation scheme in \cite{goettsch2021impact} with SNR threshold $\eta$, each UE $k$ is connected to a cluster $\Cc_k \subseteq [L]$ of RUs 
and each RU $\ell$ has a set of associated UEs $\Uc_\ell \subseteq [\Ktot]$. The UE-RU association is described by 
a bipartite graph $\Gc$ with two classes of nodes (UEs and RUs) such that the neighborhood of UE-node $k$ is $\Cc_k$ 
and the neighborhood of RU-node $\ell$ is $\Uc_\ell$. The set of edges of $\Gc$ is denoted by $\Ec$, i.e., $\Gc = \Gc([L], [\Ktot], \Ec)$. 
We assume OFDM modulation and that the channel in the time-frequency domain follows the
standard block-fading model \cite{marzetta2010noncooperative,9336188,9064545}. The channel vectors from UEs to RUs are random but constant over coherence blocks of 
$T$ signal dimensions in the time-frequency domain, of which $\taudmrs$ dimensions are used for the finite-dimensional uplink (UL) pilot signal, leading to a spectral efficiency factor of $1-\frac{\tau_p}{T}$.
The described methods are formulated for one RB, so the RB index is omitted for simplicity. 

We let $\HH \in \CC^{LM \times \Ktot}$ denote the channel matrix between all the $\Ktot$ UE antennas and all the $LM$ 
RU antennas on a given RB, formed by $M \times 1$ blocks $\hv_{\ell,k}$ in correspondence of the $M$ antennas of RU $\ell$ 
and UE $k$. 
Let $\Fm$ denote the $M \times M$ unitary DFT matrix with $(m,n)$-elements
$\left[ \Fm \right]_{m,n} = \frac{e^{-j\frac{2\pi}{M} mn}}{\sqrt{M}}$ for  $m, n  = 0,1,\ldots, M-1$, and consider the angular support set $\Sc_{\ell,k} \subseteq \{0,\ldots, M-1\}$ 
obtained according to the single ring local scattering model (see \cite{adhikary2013joint}). Then, the channel between RU $\ell$ and UE $k$ is
\begin{equation} 
	\hv_{\ell,k} = \sqrt{\frac{\beta_{\ell,k} M}{|\Sc_{\ell,k}|}}  \Fm_{\ell,k} \nuv_{\ell, k}, \label{channel_model}
\end{equation}
where, using a Matlab-like notation, $\Fm_{\ell,k} \eqdef \Fm(: , \Sc_{\ell,k})$ denotes the tall unitary matrix obtained by selecting the columns 
of $\Fm$ corresponding to the index set $\Sc_{\ell,k}$, $\beta_{\ell,k}$ is a LSFC including
pathloss, blocking effects, and shadowing, and  $\nuv_{\ell,k}$ is an $|\Sc_{\ell,k}| \times 1$ i.i.d. Gaussian vector with components 
$\sim \Cc\Nc(0,1)$. 
We focus on UL results in this work, since by duality, the UL and downlink data rates and thus the system performance are almost identical \cite{kddi_uldl_precoding}. 

\subsection{Uplink decoding under scheduling}
We assume that UL pilot allocation for channel estimation and cluster formation are carried out in each time slot after making the scheduling decisions, where the pilot allocation and cluster formation follow the semi-overloaded pilot assignment  scheme from \cite{osawa2022effective}, such that an RU can assign a pilot to multiple UEs under the condition that the UEs' channels are in orthogonal subspaces.
We let $\hh_k$ denote the $k$-th column of $\HH$. Based on the set of active UEs $\Kactset$ of cardinality $\Kact$, the columns $\hh_k$ corresponding to inactive UEs, i.e., UEs $k \in [\Ktot]: k \notin \Kactset$, contain the identically zero vector $\zerov$, since the inactive UEs do not transmit neither UL pilots nor data.
Each RU $\ell$ computes locally the channel estimates $\widehat{\hv}_{\ell,k}$ for UEs $k \in \Uc_\ell$, obtained by subspace projection channel estimation, where perfect subspace knowledge is assumed (see \cite{gottsch2022subspace} for details).

We define the {\em partial CSI} regime
where each RU $\ell$ has knowledge of the channel vector estimates $\widehat{\hv}_{\ell,k}$ for $k \in \Uc_\ell$. 
In this regime, the part of the channel matrix $\HH$ known  at the decentralized processing unit serving cluster $\Cc_k$ 
is denoted by $\widehat{\HH}(\Cc_k)$. This matrix has the same dimensions of $\HH$,  such that the $(\ell, j)$ block 
of dimension $M \times 1$  of $\widehat{\HH}(\Cc_k)$ is equal to $\widehat{\hv}_{\ell,j}$ for all  $(\ell, j) \in \Ec$, where $\ell \in \Cc_k$, and to $\zerov$ otherwise.  

Based on the channel estimates $\{\widehat{\hv}_{\ell,k}:  k \in \Uc_\ell\}$,
RU $\ell$ locally computes a unique receiver combining vector $\vv_{\ell,k}$ for each associated UE $k \in \Uc_\ell$, where a linear MMSE principle is used.
We  define the combining coefficient $w_{\ell,k}$ of RU-UE pair $(\ell,k)$ and the receiver {\em unit norm} vector $\vvv_k \in \CC^{LM \times 1}$ formed by $M \times 1$ blocks
$w_{\ell,k} \vv_{\ell,k} : \ell = 1, \ldots, L$, such that $\vv_{\ell,k} = \zerov$ if $(\ell,k)$ are not associated.
The coefficients $w_{\ell,k}$ are optimized by cluster $\Cc_k$ to maximize the UL SINR (see \cite{goettsch2021impact} for details).

\section{Uplink data transmission}
Let all active UEs transmit with the same average energy per symbol $P^{\rm ue}$, and we define the system parameter $\SNR \eqdef P^{\rm ue}/N_0$, where $N_0$ denotes the complex baseband noise power spectral density. The received $LM \times 1$ symbol vector at the $LM$ RU antennas for a single channel use of the UL is given by
\begin{equation} 
	\yy = \sqrt{\SNR} \; \HH \sss   + \zz, \label{ULchannel}
\end{equation}
where $\sss \in \CC^{K \times 1}$ is the vector
of information symbols transmitted by the UEs (zero-mean unit variance and mutually independent random variables) and 
$\zz$ is an i.i.d. noise vector with components $\sim \Cc\Nc(0,1)$.  
The goal of cluster $\Cc_k$ is to produce an effective channel observation for symbol $s_k$, the $k$-th component of the vector $\sss$, from the collectively received signal at the RUs $\ell \in \Cc_k$.  
Using the receiver vector $\vvv_k$,
the corresponding scalar combined observation for symbol $s_k$ is given by 
$ \hat{s}_k  = \vvv_k^\herm \yy. $
The instantaneous mutual information is given by 
\begin{gather}
	I_k\left( \vvv_k, \HH \right) \defines \log\left( 1 + \SINR_k \right), 
\end{gather}
where 
\begin{equation}
	\SINR_k = \frac{  |\vvv_k^\herm \hh_k|^2 }{ \SNR^{-1}  + \sum_{j \neq k} |\vvv_k^\herm \hh_j |^2 }  \label{UL-SINR-unitnorm}
\end{equation}
is the instantaneous signal to Interference plus noise ratio (SINR) value. 

\subsection{Rate allocation}
%
We consider outage rates as the effective UL data rates. Under slot-by-slot coding and decoding, the receiver can reliably decode an allocated rate $r_k$ for UE $k$ provided that no information outage occurs (see \cite{biglieri1998fading} and references therein). This condition holds if the allocated rate $r_k$ is smaller than the mutual information $I_k\left( \vvv_k, \HH \right)$.
The effective UL service rate of UE $k$ in time slot $t$, under consideration of the spectral efficiency factor $1-\frac{\tau_p}{T}$, is thus given by \cite{shirani2010mimo}
\begin{gather}
	R_k(t) = \begin{cases}
		(1 - \frac{\tau_p}{T}) R_k ,& \text{if } x_k = 1 , \\ 
		0 ,&  \text{if } x_k = 0, \end{cases} 	 \label{eq:allocated_rate}
\end{gather}
where 
\begin{gather}
	R_k \defines R_k\left( \vvv_k(t), \HH(t) \right) = r_k \times 1 \left\{ r_k < I_k \left( \vvv_k(t), \HH(t) \right) \right\}, \label{eq:def_Rk}
\end{gather}
$1 \left\{ \Ac \right\}$ is the indicator function of an event $\Ac$, $\vvv_k(t)$ and $\HH(t)$ are the realizations of $\vvv_k$ and $\HH$ in time slot $t$, and $x_k$ is the activity variable for UE $k$, which is equal to $1$ ($0$, respectively) if UE $k$ is scheduled (not active, respectively). Note that $\HH(t)$ is generated independently in each time slot according to (\ref{channel_model}). The channel estimates and $\vvv_k(t)$ for all $k \in \Kactset$ are functions of the realization of $\HH(t)$ under estimation noise.
The scheduling schemes are evaluated in terms of the UE throughput given by
\begin{align}
	\bar{R}_k = \lim\limits_{T_s \rightarrow \infty} \frac{1}{T_s} \sum_{t=1}^{T_s} R_k(t) , \label{eq:ue_throughput}
\end{align}
where $T_s$ is the total number of time slots. 

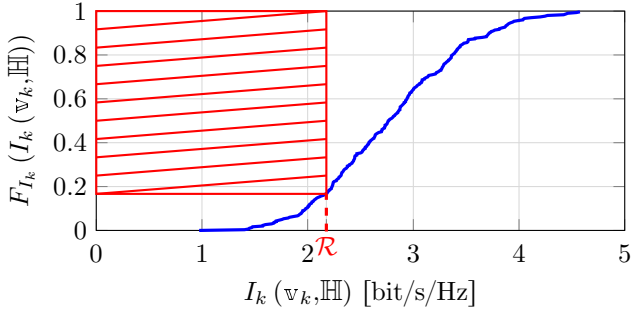
\begin{figure}[t!]
	\centering
	\begin{tikzpicture}
		\begin{axis}[
			enlarge y limits=false,clip=false,
			enlarge x limits=false,
			no markers,
			width=.95\linewidth, 
			height=4.5cm, 
			grid=major, 
			grid style={gray!30}, 
			xlabel= $I_k\left( \vvv_k { , } \HH \right)$, 
			ylabel=$F_{I_k}\left( I_k\left( \vvv_k {,} \HH \right) \right)$,
			x unit=\si{bit/s/Hz}, 
			legend style={at={(0.5,-0.2)},anchor=north}, 
			x tick label style={yshift=-2,rotate=0,anchor=north}, 
			xmin=0,
			xmax=5,
			ymin=0,
			ymax=1,
			extra x ticks={2.17707695762673},
			extra x tick labels={$\Rc$},
			extra x tick style={red, xticklabel style={yshift=3, anchor=north}}
			]
			\addplot[blue, very thick]
			table[blue,x=Var1,y=Var2,col sep=comma] {Ikdistribution.csv}; 
			\draw[red, thick] (0, 1/6) rectangle (2.17707695762673,1);
			\draw[red, thick] (0,1/6) -- (2.17707695762673,3/12);
			\draw[red, thick] (0,3/12) -- (2.17707695762673,4/12);
			\draw[red, thick] (0,4/12) -- (2.17707695762673,5/12);
			\draw[red, thick] (0,5/12) -- (2.17707695762673,6/12);
			\draw[red, thick] (0,6/12) -- (2.17707695762673,7/12);
			\draw[red, thick] (0,7/12) -- (2.17707695762673,8/12);
			\draw[red, thick] (0,8/12) -- (2.17707695762673,9/12);
			\draw[red, thick] (0,9/12) -- (2.17707695762673,10/12);
			\draw[red, thick] (0,10/12) -- (2.17707695762673,11/12);
			\draw[red, thick] (0,11/12) -- (2.17707695762673,12/12);
			\draw[red, very thick, dashed] (2.17707695762673,-.005) -- (2.17707695762673,1/6);
			\legend{}; 
		\end{axis}
	\end{tikzpicture}
	\vspace{-.2cm}
	\caption{Example empirical CDF of $I_k\left( \vvv_k, \HH \right)$ of UE $k$.}
	\label{fig:data_rate_fixedUE_cdf}
	\vspace{-.4cm}
\end{figure}

The allocated rates are computed with a semi-analytical scheme, where we assume that the decoder has knowledge of the instantaneous mutual information. Further assume that we have knowledge of a large set of realizations of $I_k \left( \vvv_k, \HH \right)$ for UE $k$. We construct an empirical cumulative distribution function (CDF) of the mutual information and, considering (\ref{eq:allocated_rate}), we can compute the allocated rate $r_k$ that maximizes the expected outage rate of UE $k$, i.e., 
\begin{align}
	r_k = \underset{ \Rc }{\arg\max}\ \ \Rc \times P_k(\Rc) , \label{eq:r_k_star}
\end{align}
where $P_k(\Rc) = \PP( \Rc  < I_k\left( \vvv_k, \HH \right) )$ and $\PP\left( \Ac \right)$ is the probability of an event $\Ac$.
This is depicted in Fig. \ref{fig:data_rate_fixedUE_cdf}, where $P_k(\Rc) = 1 - F_{I_k}(\Rc)$, and $F_{I_k}(\Rc)$ is the value of the empirical CDF of $I_k\left( \vvv_k, \HH \right)$ at $\Rc$.

 \section{System Optimization} \label{sec:system_optimization}
We consider a network in the UL with a total number of $\Ktot$ UEs, which operates at its optimal load, when serving $\Kact < \Ktot$ UEs. Further, we assume an “infinite backlog” situation, where all data to be transmitted is available at the UEs. By scheduling at most $\Kact$ UEs per time slot, a concave entry-wise non-decreasing utility function $g(\cdot)$ of the user individual ergodic rates $\bar{\Rm} = \left( \bar{R}_1, \dots, \bar{R}_{\Ktot} \right)$ shall be maximized. The problem to be solved is 
\begin{eqnarray}
	& \text{maximize} & g(\bar{\Rm}) , \ \ \ \ \text{subject to } \bar{\Rm} \in \Rs ,  \label{eq:max_utility} 
\end{eqnarray}
where $\Rs$ is the achievable ergodic rate region \cite{georgiadis2006resource}. The solution of (\ref{eq:max_utility}) is denoted by $\bar{\Rm}^\star$, which is generally very hard to find, since $\Rs$ is not characterized easily \cite{shirani2010mimo}. 
However, we can use the framework of \cite{georgiadis2006resource} to find a scheduling scheme that approximates $\bar{\Rm}^\star$. For this aim, we use ``virtual queues'' driven by ``virtual arrival processes'', such that the arrival rates approximate $\bar{\Rm}^\star$ (see e.g. \cite{georgiadis2006resource, shirani2010mimo}). 
 
Specifically, in time slot $t$, the arrival processes are given by $A_k(t) = a_k$, where $\av = (a_1, \dots, a_{\Ktot})$ is the solution to the convex optimization problem
\begin{equation}
	\begin{array}{l l}
		 \underset{\av}{\text{maximize}} & \ Vg(\av) - \sum_{k \in [\Ktot]} Q_k\left( t \right) a_k  \\
		 \text{subject to} & \ 0 \leq a_k \leq A_{\rm max}, \ \ \forall k \in [\Ktot]. 
	\end{array} \label{eq:opt_ak_general}
\end{equation}
where $V, \ A_{\rm max}$ are suitably chosen constant parameters that determine the behavior of the algorithm, and $Q_k\left( t \right)$ is the virtual queue of user $k$ in time slot $t$. 
The virtual queues are updated in each time slot after computing the arrivals $A_k(t)$ and service rates $R_k(t)$ according to 
\begin{align}
	Q_k(t+1) = \max\left\{ Q_k(t) - R_k(t), 0 \right\} + A_k(t). \label{eq:queue_update}
\end{align}

\subsection{Uplink scheduling scheme}
In time slot $t$, we solve the optimization problem (\ref{eq:opt_ak_general}) yielding the arrivals $A_k(t)$, and use the virtual queues $Q_k(t)$ as weights in the scheduling problem. Having defined $\Kact$ as the desired number of simultaneously active UEs, we solve 
\begin{subequations}
	\begin{eqnarray}
		& \underset{\xv}{\rm maximize} & \sum_{k \in [\Ktot]} Q_k(t) \EE\left[ R_k \right] x_k  \\
		& \text{subject to} & \sum_{k \in [\Ktot]} x_k \leq \Kact ,  \\
		&  & x_k \in \left\{ 0, 1 \right\} , 
	\end{eqnarray}
\end{subequations}
where $\EE \left[ R_k \right] = r_k \times P_k(r_k) $ is the expected service rate of UE $k$ and $r_k$ is given by (\ref{eq:r_k_star}).
Since $Q_k(t)$ and $\EE\left[ R_k \right]$ are known, this problem can be solved by simply scheduling the $\Kact$ UEs with the largest product $Q_k(t) \EE\left[ R_k \right] $.
This results in the service rates $R_k(t)$ given by (\ref{eq:allocated_rate}) and (\ref{eq:def_Rk}).
The virtual queues are then updated with the computed arrivals $A_k(t)$ and the service rates $R_k(t)$ according to (\ref{eq:queue_update}).

\subsection{Proportional fairness and hard fairness scheduling}
We consider PFS and HFS, leading to different solutions to the optimization problem (\ref{eq:opt_ak_general}).
In case of considering PFS, we have $g(\av) = \sum_{k \in [\Ktot]} \log a_k$ and the convex optimization problem
\begin{subequations}
	\begin{eqnarray}
		& \underset{\av}{\text{maximize}} & V \sum_{k \in [\Ktot]} \log a_k - \sum_{k \in [\Ktot]} Q_k\left( t \right) a_k \label{schedule_opt_pf} \\
		& \text{subject to} & 0 \leq a_k \leq A_{\rm max}, \ \ \forall k ,
	\end{eqnarray}
\end{subequations}
which can be solved using Lagrange-KKT conditions yielding the arrivals
	\begin{align}
		a_k = \min \left\{ \frac{V}{Q_k(t)}, A_{\rm max} \right\} . \label{eq:a_k_pfs}
	\end{align}
For HFS, we have $g(\av) = \underset{k \in [\Ktot]}{\min} \ a_k$ and the convex optimization problem
\begin{subequations}
	\begin{eqnarray}
		& \underset{\av}{\text{maximize}} & V \kappa - \sum_{k \in [\Ktot]} Q_k\left( t \right) a_k \\
		& \text{subject to} & 0 \leq \kappa \leq a_k \leq A_{\rm max}, \ \ \forall k \in [\Ktot]  \label{eq:max_min_a_k_constraint_1} ,
	\end{eqnarray}
\end{subequations}
where $\kappa$ is an auxiliary variable.
In this case, the solution is given by  \cite{shirani2010mimo}
\begin{align}
	a_k = \begin{cases}  A_{\rm max} , & \text{if } V > \sum_{k \in [\Ktot]} Q_k(t), \\ 
		0 , & \text{else. }  \end{cases} \label{eq:a_k_hf}
\end{align}

\subsection{Online rate adaptation}
The allocated rates $r_k$ for all $k \in [\Ktot]$ are learned online. We consider a decoder that has knowledge of the instantaneous mutual information. 
For each UE $k$, we consider a memory of the last $N$ samples of the mutual information. 
In each time slot $t$, in which UE $k$ is scheduled, the mutual information $I_k \left( \vvv_k(t), \HH(t) \right)$ for UE $k$ is saved to the memory and replaces the oldest sample. By that scheme, we maintain a memory of the latest $N$ values of $I_k \left( \vvv_k, \HH \right)$ and continuously update the allocated rates with (\ref{eq:r_k_star}).
Note that the allocated rates are initialized by a ``start-up'' phase consisting of $N_{\rm init}$ time slots, where in each of the $N_{\rm init}$ time slots $\Kact$ out of the $\Ktot$ UEs are randomly selected to be active.
For each UE, we save the samples of the mutual information experienced during the start-up phase and compute the initial allocated rate with (\ref{eq:r_k_star})\footnote{We use this constructed ``start-up'' phase to achieve an allocated rate for each UE. In practical systems, this may be done on a per UE basis, i.e., when a UE joins the system, by some admission control scheme.}.

\section{Numerical evaluations and outlook}
\begin{figure}[t!]
	\centerline{\includegraphics[width=.5\linewidth]{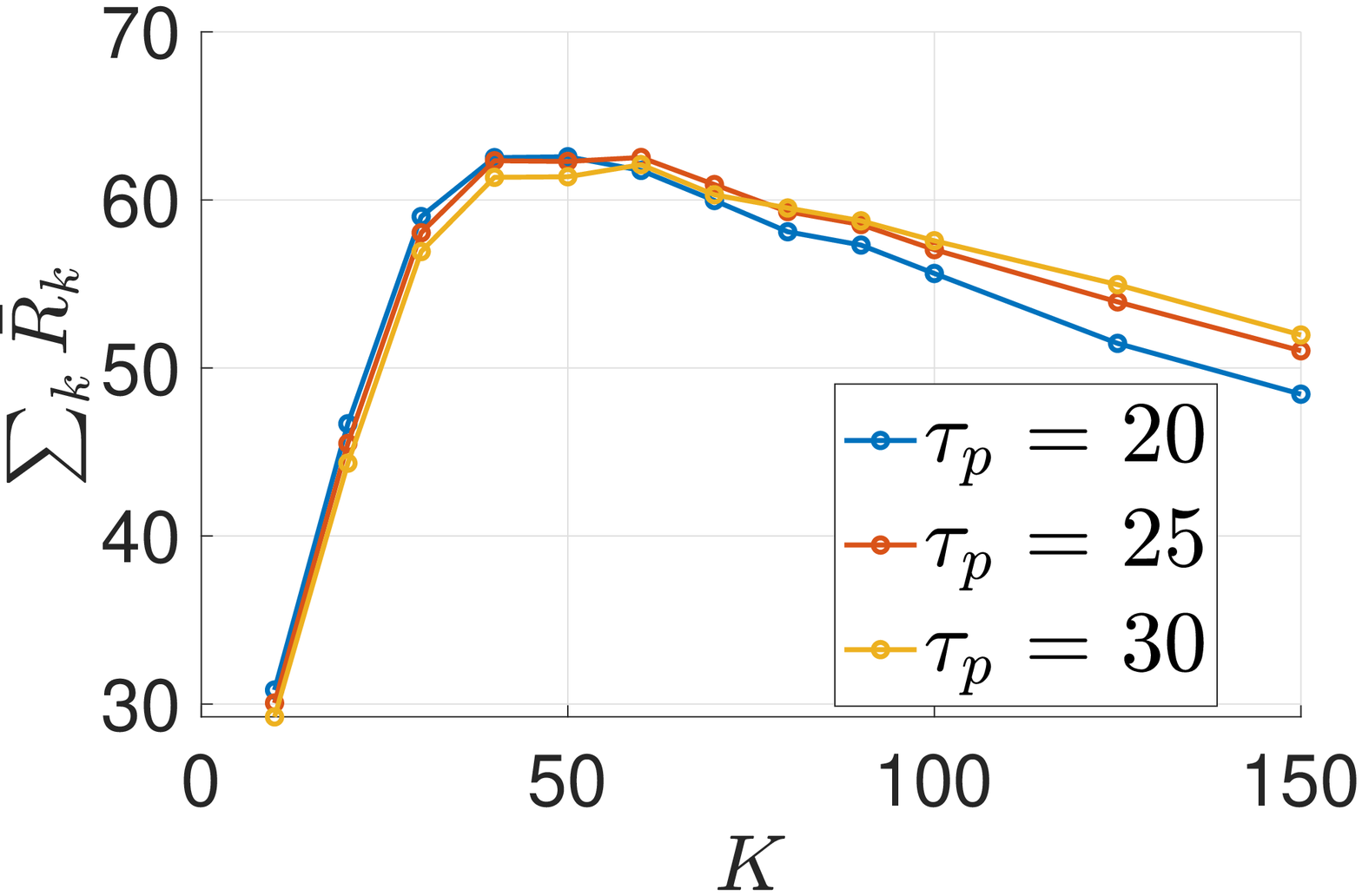} 
	\includegraphics[width=.5\linewidth]{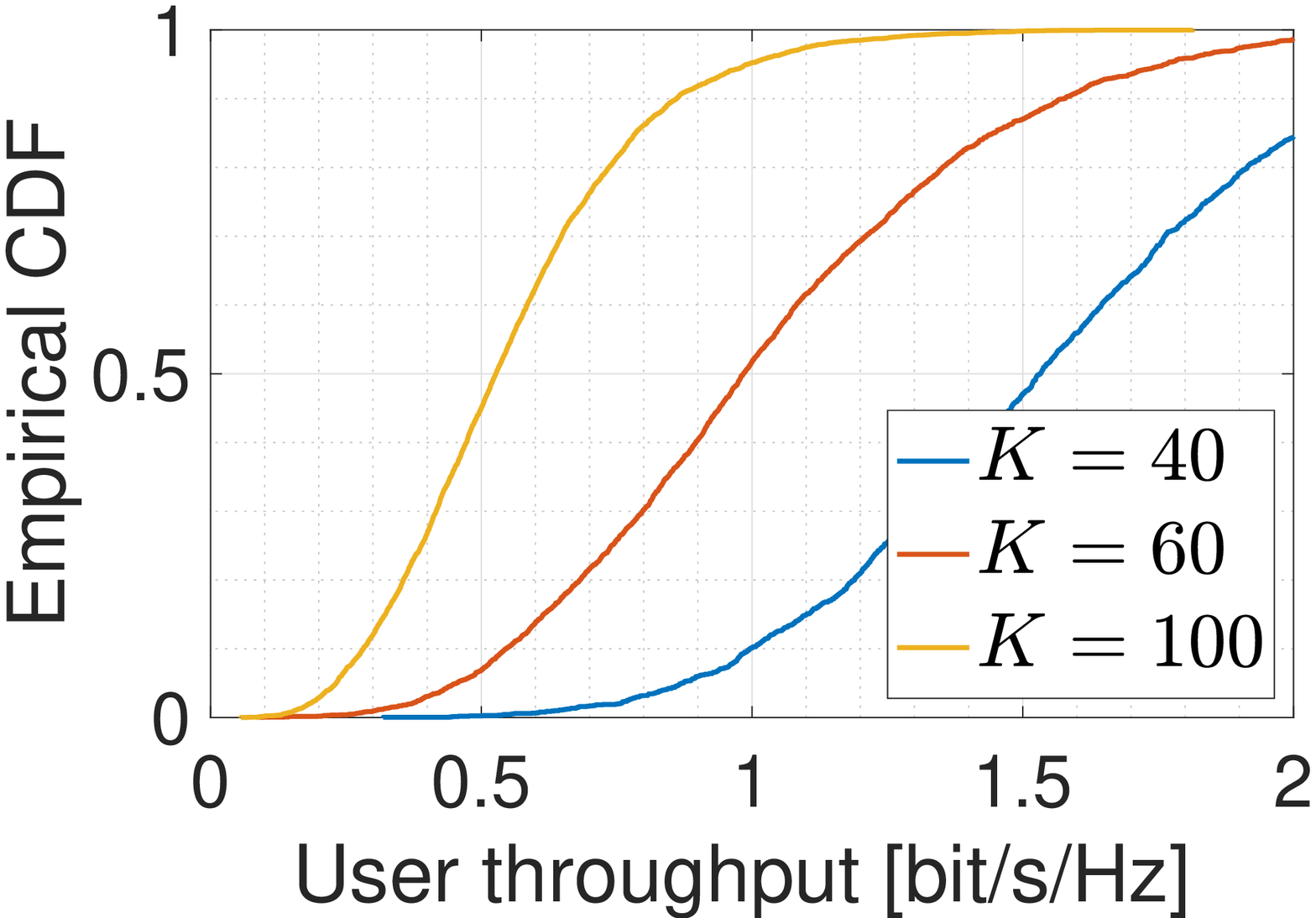}  }
	\vspace{-.25cm}
	\caption{Sum throughput vs. $\Ktot$ for different $\tau_p$ (left). The empirical CDF of the user throughput for $K=\left\{40, 60, 100\right\}$ (right).}
	\label{fig_sum_se_vs_K_L12}
	\vspace{-.45cm}
\end{figure}
\begin{figure}[t!]
	\centerline{\includegraphics[width=.5\linewidth]{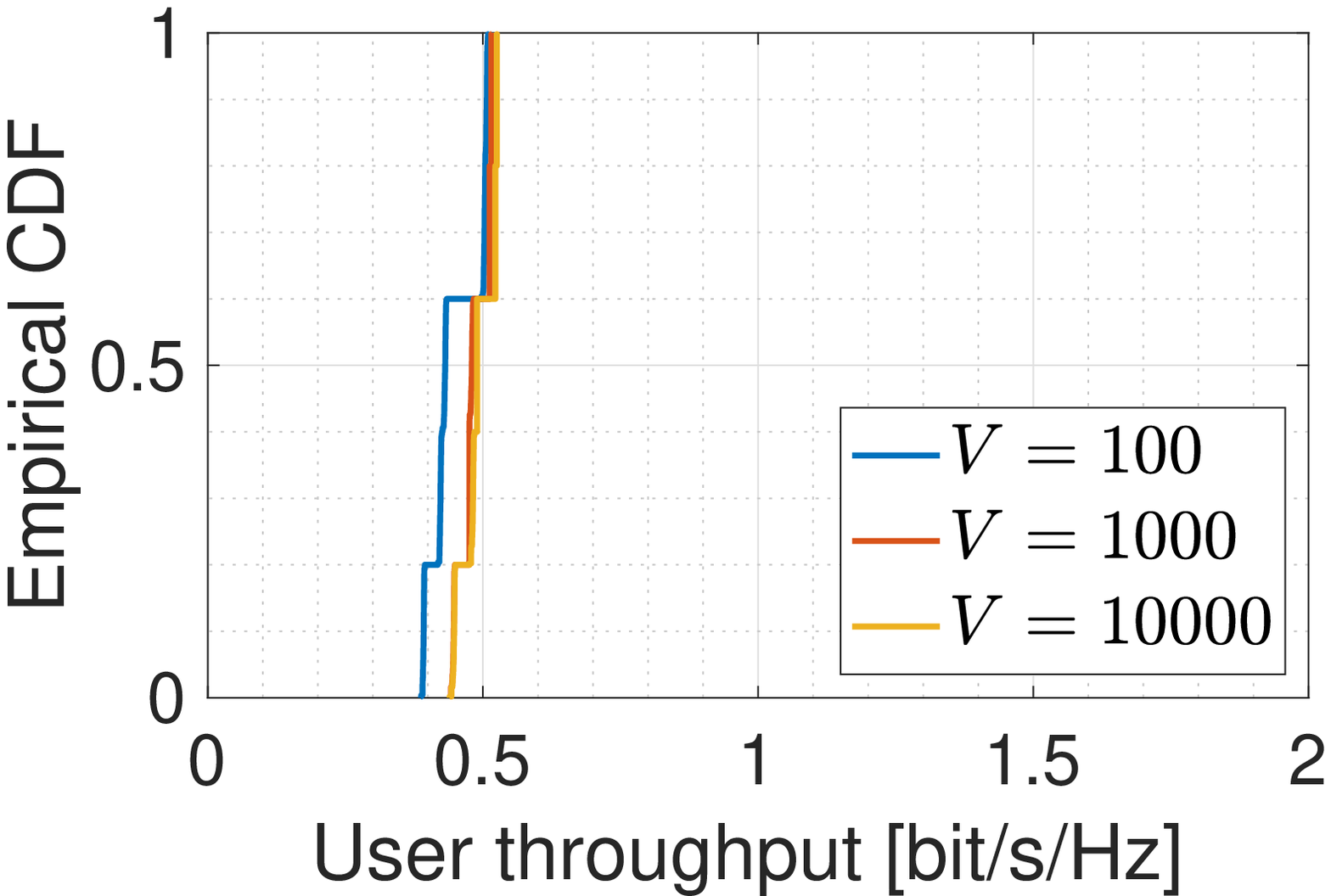} \includegraphics[width=.5\linewidth]{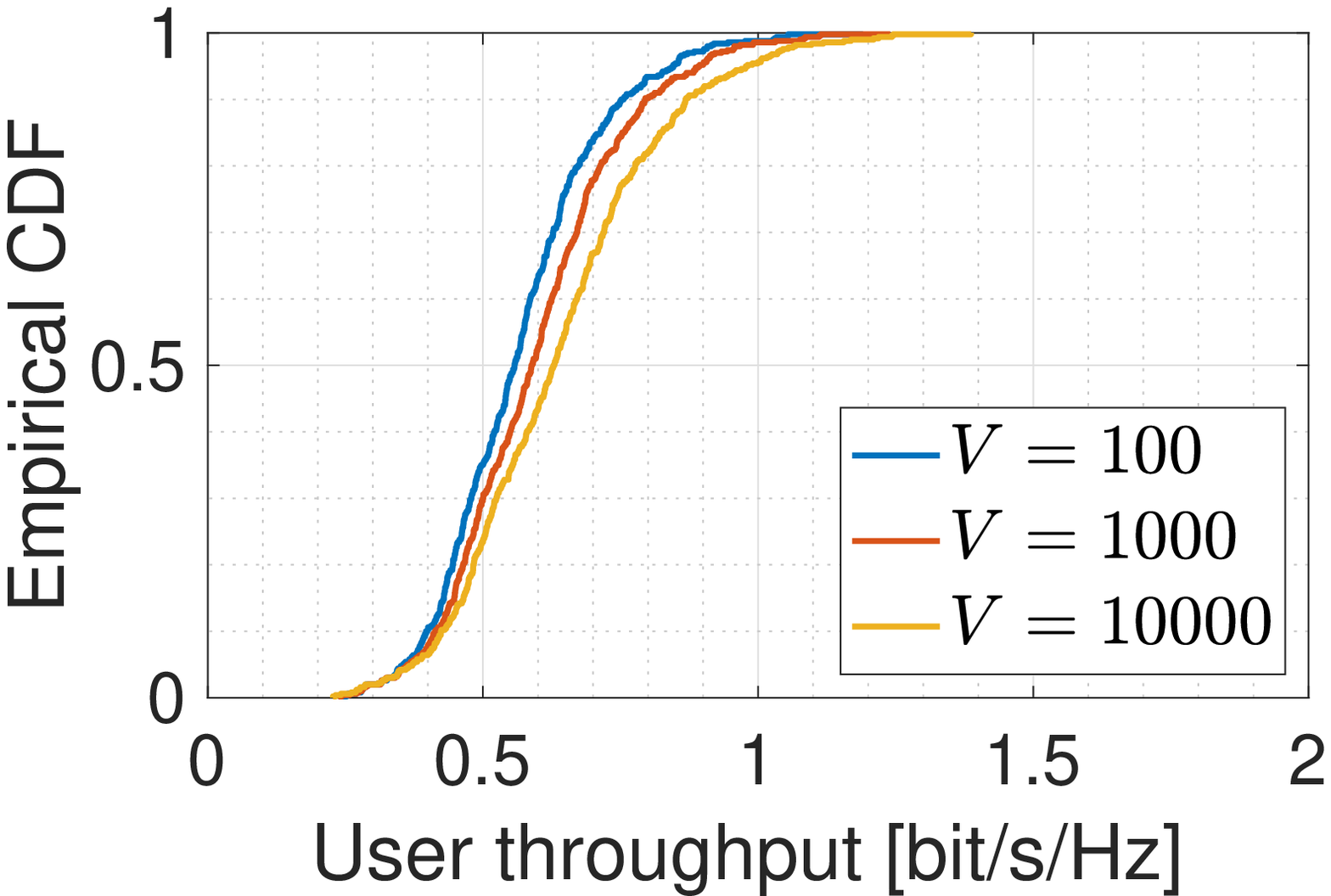}}
	\vspace{-.25cm}
	\caption{HFS (left) and PFS (right) with different $V$.}
	\label{fig:pf_and_hf_cdf_plot}
	\vspace{-.6cm}
\end{figure}
We consider a dense cell-free network spanning an area of $50 \times 50$ $\text{m}^2$ with a torus topology to avoid boundary effects, containing $L=12$ RUs, each with $M=8$ antennas. 
A bandwidth of $W= 10$ MHz and noise with power spectral density of $N_0 = -174$ dBm/Hz is assumed. 
The angular support $\Sc_{\ell,k}$ contains the DFT quantized angles (multiples of $2\pi/M$) falling inside an interval of length $\Delta$ placed symmetrically around the direction joining UE $k$ and RU $\ell$. We use $\Delta = \pi/8$ and the maximum cluster size $Q=10$ (RUs serving one UE) in the simulations. The SNR threshold $\eta=1$ makes sure that an RU-UE association can only be established, when $\beta_{\ell,k} \geq \frac{\eta}{M \SNR } $.
The UL energy per symbol is chosen such that $\bar{\beta} M \SNR = 1$ (i.e., 0 dB), when the expected pathloss $\bar{\beta}$ with respect to LOS and NLOS is calculated for distance $3 d_L$, where $d_L = \sqrt{\frac{A}{\pi L}}$ is the radius of a disk of area equal to $A/L$. The actual (physical) UE transmit power is obtained as $P_{\rm tx}^{\rm ue} = P^{\rm ue}W$. This leads to a certain level of overlap of the RUs' coverage areas, such that each UE is likely to be associated to several RUs. The UEs are randomly dropped in the network area, while the RUs are placed on $3\times 4$ rectangular grid. The online rate adaptation is carried out for all schemes with  $N_{\rm init} = 500$ and $N = 100$, and we consider RBs of dimension $T = 200$ symbols.

\subsection{Optimal user load}
We start our evaluations by finding the optimal user load in terms of the sum throughput $\sum_{k=1}^{\Ktot} \bar{R}_k$. For each set of parameters we generated 50 independent layouts (random uniform placement of UEs), 
and for each  layout we simulated $T_s = 1000$ time slots, in which all $\Ktot$ UEs are active.
The sum throughput is maximized with approximately $\Ktot \in \left[40, 60\right]$ simultaneously active UEs, as depicted in Fig. \ref{fig_sum_se_vs_K_L12}, where the average sum throughput of all topologies is shown. With small $\Ktot$, the network does not operate at its full capability and thus a low sum throughput is achieved.
For heavily loaded networks with $\Ktot \gg \frac{ML}{2}$, the sum throughput decreases again due to the high interference.
We want to operate the network close to its maximum throughput, but not at ``full load'', where a not negligible number of UEs experiences a low throughput, as Fig. \ref{fig_sum_se_vs_K_L12} shows for $\Ktot = \{60, 100\}$. We thus focus on having $\Kact = 40$ simultaneously active UEs per time slot and $\tau_p = 20$ in the following for the fairness scheduling problem.
%

\subsection{Utility optimization}
We consider the same network as before with a total number of $\Ktot =100$ UEs, where we schedule $\Kact = 40$ UEs per time slot. The proposed method using virtual arrivals and queues for HFS and PFS are compared with random and round-robin scheduling, as well as with a max-sum-rate scheduler (``Max. sum rate'').
With random scheduling, we randomly pick $\Kact$ UEs per time slot, independent of the previous scheduling decisions. With round-robin scheduling, we sort the UEs by their index and schedule them in a subsequent way, such that, e.g., in the first three time slots the following UEs are active: $t=1:\{1, \dots, 40\}, \ t=2:\{41, \dots, 80\}, \ t=3:\{81, \dots, 20\}$. The max-sum-rate scheduler selects in each time slot the $\Kact$ UEs to be active that have the largest expected service rate.
We consider five different topologies, where the UEs are randomly placed. The empirical CDF plots contain the throughput of the UEs from all topologies. The average sum log user throughput is given by computing  $\sum_{k \in [\Ktot]} \log \left( \bar{R}_k \right)$ for each topology separately, and then taking the average over the topologies. The simulations are stopped, when the queues of all UEs reach a steady state.

Fig. \ref{fig:pf_and_hf_cdf_plot} shows the empirical CDF of the user throughput with HFS and PFS for different $V$. With HFS, $V = \{1000,10000\}$ yield very similar results, achieving a significant improvement of the throughput compared to $V = 100$. 
With PFS, increasing $V$ leads to a larger throughput of most UEs. 
Specifically, the throughput of UEs in the upper part of the empirical CDF is improved, while the throughput of the UEs in the lower part is quite similar. Compared to the results in Fig. \ref{fig_sum_se_vs_K_L12}, where all $\Ktot=100$ UEs are active in each time slot, the performance of UEs with low throughput is significantly improved. The network with $\Ktot = 100$ simultaneously active UEs is ``congested'', leading to a not negligible fraction of UEs suffering from high interference.
For, example the 10th percentile throughput is below $0.3 \text{ bit/s/Hz}$ with all $\Ktot$ UEs active in all time slots, compared to approximately $0.45 \text{ bit/s/Hz}$ achieved by both HFS and PFS with $V=10000$. This improvement leads to more fairness, motivating the proposed approaches, but comes at the cost of a decreased performance of UEs with a large throughput, especially when HFS is employed.

\begin{figure}[t!]
	\centerline{\includegraphics[width=.5\linewidth, trim=5 5 2 7, clip]{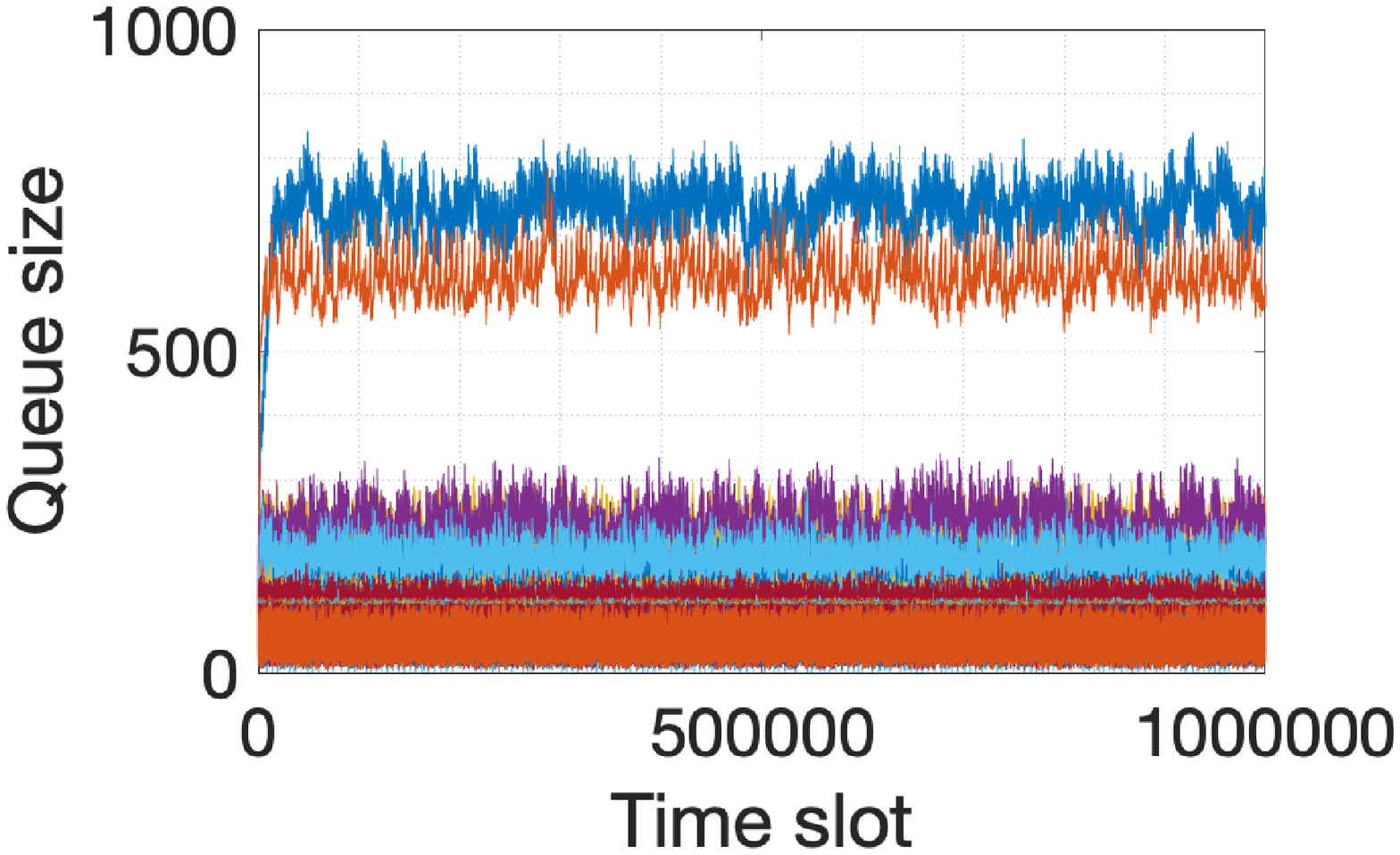} \includegraphics[width=.5\linewidth, trim=5 5 2 7, clip]{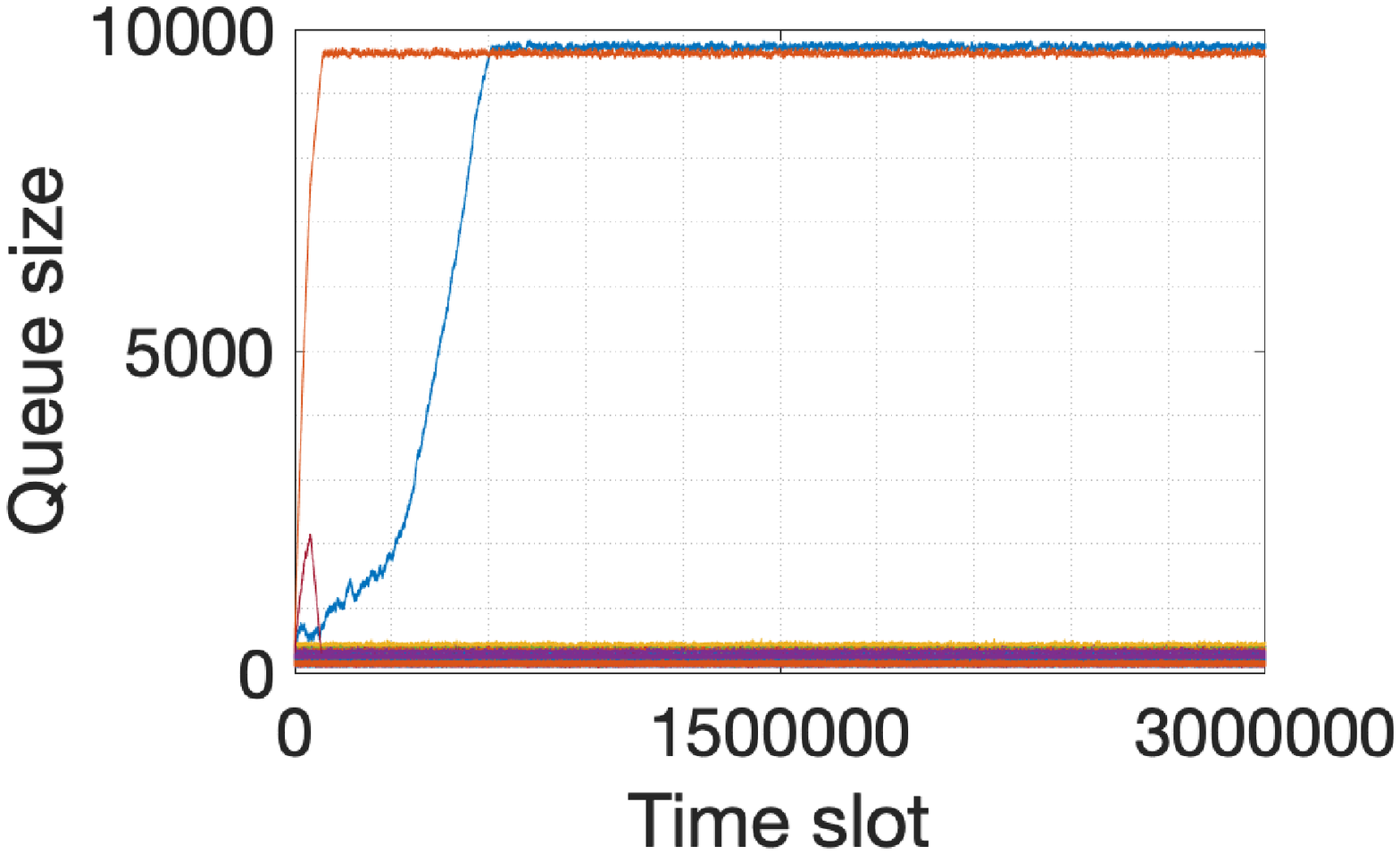} }
	\vspace{-.2cm}
	\caption{The longest queues using HFS with $V=1000$ (left) and $V=10000$ (right).}
	\label{fig:pf_and_hf_cdf_queues}
	\vspace{-.4cm}
\end{figure}

The better performance with larger $V$ for both HFS and PFS comes at the cost of more congestion of the queues, such that more time slots are required until the queues of all UEs reach a steady state. Both observations are depicted in Fig. \ref{fig:pf_and_hf_cdf_queues} and can be theoretically justified (see \cite{georgiadis2006resource}). The evolution of the queues for HFS  and $V= \{1000, 10000\}$ is shown in Fig. \ref{fig:pf_and_hf_cdf_queues}, where from all topologies the hundred queues are shown, which have the largest values in the last slot. A remarkable observation is that very few UEs have a long queue in most time slots, while the queues of the remaining UEs reach a steady state after relatively few time slots. A similar behavior is observed with PFS, where the relative difference of the queue sizes is smaller. 

\begin{figure}[t!]
	\centerline{\includegraphics[width=.5\linewidth]{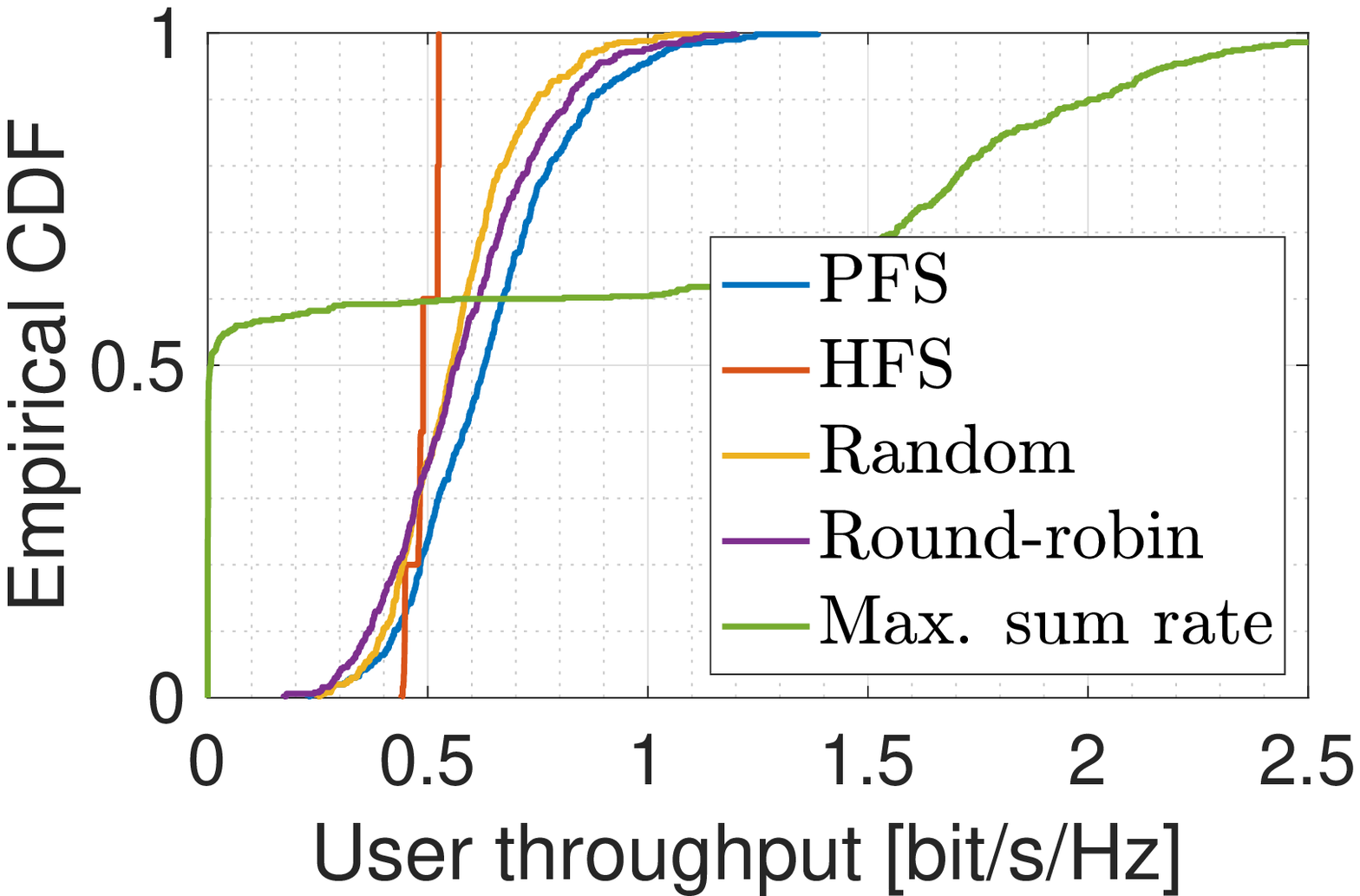} \includegraphics[width=.5\linewidth]{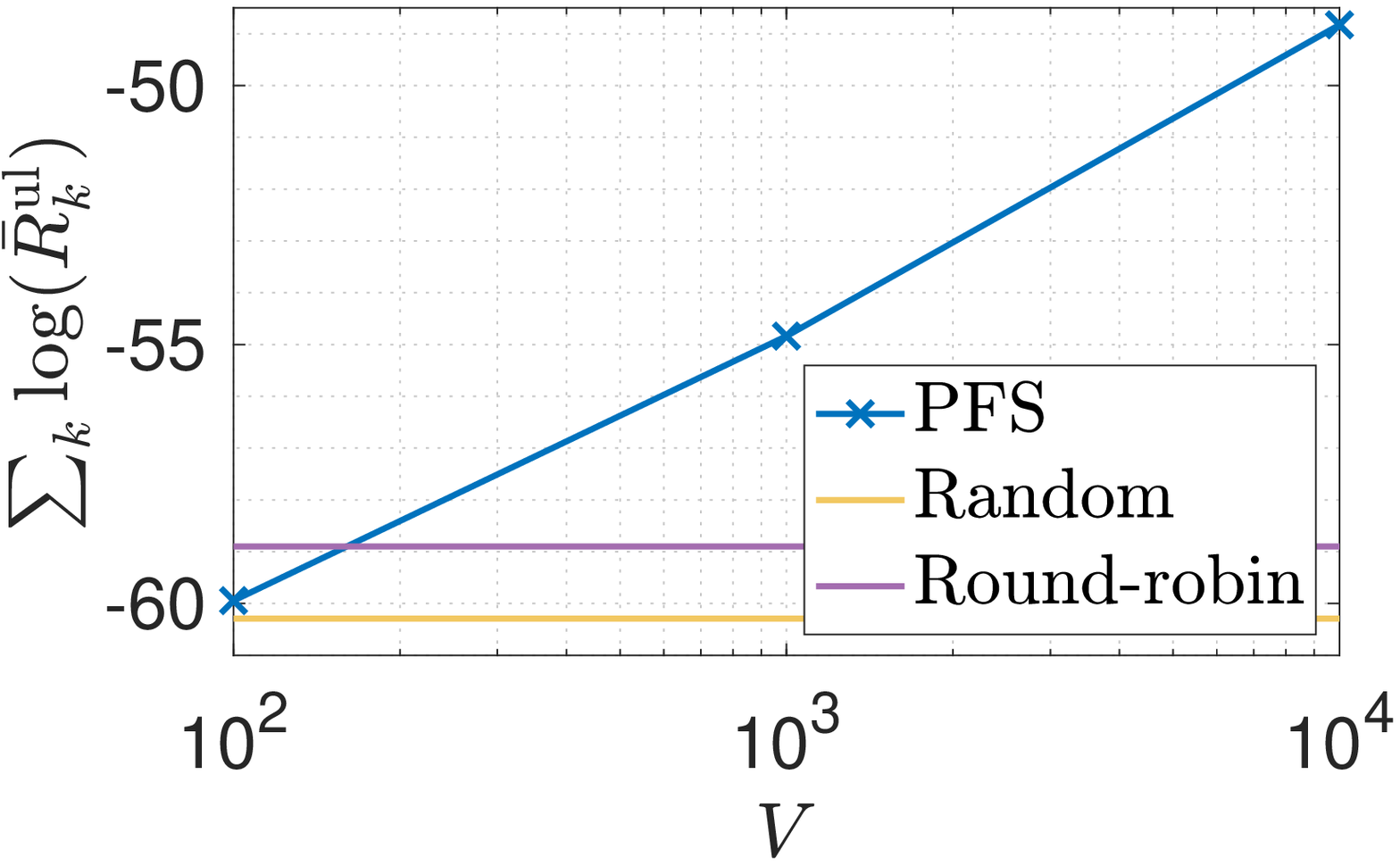} }
	\vspace{-.25cm}
	\caption{The empirical CDF of the user throughput, where $V=10000$ for HFS and PFS (left). The average sum log user throughput vs. $V$ for  PFS compared to random and round-robin scheduling (right).}
	\label{fig:pf_vs_hf_vs_rnd_vs_roro_cdf}
	\vspace{-.6cm}
\end{figure}

The performance of the proposed HFS and PFS schemes are compared to random, round-robin and max-sum-rate scheduling in Fig. \ref{fig:pf_vs_hf_vs_rnd_vs_roro_cdf}. The corresponding utility function is improved by HFS and PFS, respectively, compared to the other schemes. Random and round-robin scheduling achieve very similar results due to the random locations of the scheduled UEs.
If a network aims at maximizing the sum throughput, the system becomes \textit{very unfair}, with a large fraction of UEs with zero throughput. Note that the networks in previous works, which consider all UEs to be active with $K < LM$, translate in a very unfair system, when the actual number of UEs in the network is much larger than the $K$ considered. These systems would not serve the UEs above the limit of the desired $K$, i.e., a behavior similar to ``Max. sum rate'' in Fig. \ref{fig:pf_vs_hf_vs_rnd_vs_roro_cdf}, which is absolutely unacceptable for a real-world network.

\subsection{Concluding remarks}
The proposed methods can increase fairness among users in a cell-free system compared to random, round-robin and max-sum-rate scheduling, where the total number of UEs in the network is larger than the optimal user load. In addition, a fairer throughput distribution is achieved compared to the approach of serving all $\Ktot$ UEs simultaneously in all time slots, which leads to a congested network. The user queues with HFS suggest that a user admission and rejection scheme may improve the overall system performance, since only a very small fraction of UEs has long queues, which may prevent the system from serving the other UEs more frequently. A user admission and rejection scheme is considered future work. Distributed approaches, and a less frequent realization of pilot allocation and cluster formation are other future research points.


\bibliography{IEEEabrv,asilomar22-paper}

\end{document}